\pdfoutput=1

\documentclass[review]{elsarticle}

\usepackage{hyperref}
\usepackage{mathtools}
\usepackage{siunitx}
\usepackage{amsmath}
\usepackage{algorithm}
\usepackage{algpseudocode}
\usepackage{varwidth}
\usepackage[margin=1.5in]{geometry}

\DeclareMathOperator{\sgn}{sgn}

\journal{Computer Methods in Applied Mechanics and Engineering}

\begin{document}

\begin{frontmatter}

\title{Riemann solvers and pressure gradients in Godunov-type schemes for variable density incompressible flows}

\author{Shannon Leakey} \ead{s.c.leakey2@newcastle.ac.uk}
\author{Vassilis Glenis\fnref{myfootnote}} \ead{vassilis.glenis@newcastle.ac.uk}
\author{Caspar J.M. Hewett} \ead{caspar.hewett@newcastle.ac.uk}
\address{School of Engineering, Newcastle University, Newcastle upon Tyne, United Kingdom}
\fntext[myfootnote]{Corresponding author}

\begin{abstract}
Variable density incompressible flows are governed by parabolic equations.
The artificial compressibility method makes these equations hyperbolic-type, which means that they can be solved using techniques developed for compressible flows, such as Godunov-type schemes.
While the artificial compressibility method is well-established, its application to variable density flows has been largely neglected in the literature.
This paper harnesses recent advances in the wider field by applying a more robust Riemann solver and a more easily parallelisable time discretisation to the variable density equations than previously.
We also develop a new method for calculating the pressure gradient as part of the second-order reconstruction step.
Based on a rearrangement of the momentum equation and an exploitation of the other gradients and source terms, the new pressure gradient calculation automatically captures the pressure gradient discontinuity at the free surface.
Benchmark tests demonstrate the improvements gained by this robust Riemann solver and new pressure gradient calculation.
\end{abstract}


\end{frontmatter}

\section{Introduction}
The Navier-Stokes equations do not include an equation for pressure.
For incompressible flows, numerical methods must find a way around this problem to update the pressure field.
One approach is artificial compressibility, which adds a pseudo-time derivative of pressure to the incompressibility constraint.
This makes the governing equations hyperbolic, allowing the solution to advance in pseudo-time until it converges, at which point the pseudo-time derivative disappears and the original equations are satisfied \cite{Drikakis2005,Toro2009,Kwak2011}.
As the equations are hyperbolic, they can be solved with methods designed for compressible flows, such as Godunov-type schemes.
Chorin \cite{Chorin1997} created artificial compressibility for steady-state cases, but it has since been generalised to transient cases using dual time stepping, that is, driving the solution to the incompressible limit at each time step.

Artificial compressibility has a key computational advantage compared to the most popular option for pressure-velocity coupling, the predictor-corrector methods based on matrix inversion \cite{Patankar1972,Issa1986}.
The artificial compressibility method is easier to parallelise, especially if the pseudo-time stepping is explicit \cite{Nourgaliev2004,Loppi2019,Vermeire2019}.
Even for implicit schemes, Hodges \cite[p.5]{Hodges2020} notes more generally that artificial compressibility ``replaces the global solution of an elliptic matrix inversion with local hyperbolic wave propagation,'' making it simpler, and therefore more efficient, for processors to communicate with each other.
Consequently, the artificial compressibility method has much potential for application on modern hardware architectures \cite{Loppi2019, Vermeire2019}.

Despite this potential for parallelisation, the artificial compressibility method has not been thoroughly developed for free-surface flows \cite{Shapiro2005}, which are of much interest in engineering.
For example, waves on ship hulls \cite{Pan2000}, droplet splashes \cite{Bhat2019}, and sloshing in containers \cite{Kelecy1997} are all types of free-surface flows.
In the artificial compressibility framework, researchers have modelled free-surface flows by coupling the artificial compressibility equations with the volume of fluid (VOF) equation \cite{Pan2000,Bhat2019,Bhat2019a,Ntouras2020} and level-set equation \cite{Nourgaliev2004}.
However, the most developed method is to couple the equations with a transport equation for density, that is, to solve the variable-density artificial-compressibility equations, and to mark the location of the free surface by the density discontinuity.
Kelecy and Pletcher \cite{Kelecy1997} pioneered this method in a Godunov-type scheme, Qian et al.\ \cite{Qian2006} added the Cartesian cut-cell technique to simulate moving bodies, and Bassi et al.\ \cite{Bassi2018} used the method as part of a discontinuous Galerkin scheme, also deriving shock and rarefaction relations for the equations.
Even though this is the most developed free-surface method in the artificial compressibility framework, these three papers \cite{Kelecy1997,Qian2006,Bassi2018} were spaced out over several decades.
Before the method can be used practically and with confidence, it needs to be systematically explored, with advances from the wider computational fluid dynamics (CFD) literature incorporated, and some specific problems solved.

In this study, we develop a Godunov-type scheme for the variable-density artificial-compressibility equations, as in \cite{Kelecy1997,Qian2006}.
The scheme is general, theoretically suitable for viscous flows on 3D unstructured meshes.
Here we implement it for inviscid flows on 2D Cartesian meshes, and demonstrate its effectiveness against four sets of benchmark tests.
Before setting out the specific objectives of this study, first note that Godunov-type schemes calculate numerical fluxes at cell interfaces with Riemann solvers.
One of the most common methods to make Godunov-type schemes second-order accurate in space is MUSCL (Monotonic Upstream-centered Scheme for Conservation Laws) reconstruction \cite{vanLeer1976}.
The focus of this study is threefold: robust Riemann solvers, pressure gradients for MUSCL reconstruction, and ease of parallelisation.

\paragraph{Riemann solvers}
In the variable-density artificial-compressibility equations, the density discontinuity is marked by a contact wave.
Consequently, the free surface can be captured by the contact wave in a Riemann solver, without resorting to surface-capturing or surface-tracking methods such as VOF.
This means that the robustness of the Riemann solver is paramount.
In this study, we apply for the first time the Osher Riemann solver of Dumbser and Toro \cite{Dumbser2011} to the variable-density artificial-compressibility equations.
The results are compared with the Riemann solver of Bassi et al.\ \cite{Bassi2018}, exact under certain wave configurations, and the Roe Riemann solver implemented by Kelecy and Pletcher \cite{Kelecy1997} and Qian et al.\ \cite{Qian2006}.

\paragraph{Pressure gradients}
Without special treatment, standard MUSCL reconstruction will not account for the impact of a density discontinuity on the pressure gradient.
This can result in an imbalance of forces at the free surface and, in turn, either parasitic velocities \cite{DeGroot2011,Efremov2017,Hashemi2020} or erroneous compression at what should be the incompressible limit.
As will be shown later, the compression can resemble Gibbs-type oscillations, but it is a distinct problem of its own that persists even with the most severe of limiters.
For the variable-density artificial-compressibility equations, only Qian et al.\ \cite{Qian2006} have addressed this problem and only in the hydrostatic case; they did not consider how there can still be a force imbalance in the absence of gravity.
However, it has been investigated more comprehensively in the wider CFD literature, both for hydrostatic pressure \cite{Zingale2002,Botta2004,Krause2019,Fuchs2010} and non-hydrostatic pressure \cite{Ntouras2020,Queutey2007,Kruisbrink2018,Vukcevic2018}.
In this study, we introduce a new method for calculating the pressure gradient that has not been considered before either in the artificial compressibility literature or the more general CFD literature.
Based on a rearrangement of the momentum equation and exploitation of the other total variation diminishing (TVD) gradients and source terms, it captures the pressure gradient discontinuity at the free surface automatically.

\paragraph{Ease of parallelisation}
Kelecy and Pletcher \cite{Kelecy1997}, Qian et al.\ \cite{Qian2006}, and Bassi et al.\ \cite{Bassi2018} all used implicit schemes.
However, this fails to take advantage of one of the key advantages of artificial compressibility.
If the pseudo-time stepping is explicit, then it is relatively easy to parallelise on modern hardware architectures \cite{Nourgaliev2004,Loppi2019,Vermeire2019}.
In this study, we use the explicit Runge-Kutta scheme of Vermeire et al.\ \cite{Vermeire2019} to advance in pseudo-time.
This is a first for the variable-density artificial-compressibility equations.

\section{Numerical method}
\subsection{Governing equations}
\subsubsection{Artificial compressibility}
Variable-density incompressible flows are governed by the conservative equations
\begin{gather}
    \frac{\partial \rho}{\partial t} \label{eq:NS_start}
    + \nabla \cdot (\rho \mathbf{u})
    = 0
    \\
    \frac{\partial}{\partial t} (\rho \mathbf{u})
    + \nabla \cdot (\rho \mathbf{u} \otimes \mathbf{u})
    = -\nabla p 
    + \nabla \cdot \left(\mu \left(\nabla \mathbf{u} + \nabla \mathbf{u} ^T\right)\right)
    + \rho \mathbf{g}
    \\
    \nabla \cdot \mathbf{u} = 0 \label{eq:NS_end}
\end{gather}
where $t$ is (real-)time, $\rho$ is density, $\mathbf{u}$ is velocity, $p$ is pressure, $\mu$ is dynamic viscosity, and $\mathbf{g}$ is gravity \cite{Kelecy1997,Qian2006,Bassi2018}.
Pseudo-time derivatives are added to all three equations \eqref{eq:NS_start}--\eqref{eq:NS_end} to get
\begin{gather}
    \frac{\partial \rho}{\partial \tau} \label{eq:AC_start}
    + \frac{\partial \rho}{\partial t}
    + \nabla \cdot (\rho \mathbf{u})
    = 0
    \\
    \frac{\partial}{\partial \tau} (\rho \mathbf{u})
    + \frac{\partial}{\partial t} (\rho \mathbf{u})
    + \nabla \cdot (\rho \mathbf{u} \otimes \mathbf{u})
    = -\nabla p 
    + \nabla \cdot \left(\mu \left(\nabla \mathbf{u} + \nabla \mathbf{u} ^T\right)\right)
    + \rho \mathbf{g}
    \\
    \frac{1}{\beta} \frac{\partial p}{\partial \tau} +
    \nabla \cdot \mathbf{u} = 0 \label{eq:AC_end}
\end{gather}
where $\tau$ is pseudo-time and $\beta$ is the artificial compressibility coefficient.

\subsubsection{Compact form}
The governing equations \eqref{eq:AC_start}--\eqref{eq:AC_end} can be written in the compact form
\begin{equation} \label{eq:compact_form}
    \frac{\partial \mathbf{Q}}{\partial \tau}
    + \mathbf{I}_C \frac{\partial \mathbf{Q}}{\partial t}
    + \nabla \cdot \mathbf{F}
    = \mathbf{B}
\end{equation}
where the conservative state vector is
\begin{equation}
    \mathbf{Q}
    =
    \begin{bmatrix}
        \rho \\ \rho u \\ \rho v \\ \rho w \\ p/\beta
    \end{bmatrix},
\end{equation}
the cancellation matrix \cite{Vermeire2019} is
\begin{equation}
    \mathbf{I}_C
    =
    \begin{bmatrix}
        1 & & & & \\
        & 1 & & & \\
        & & 1 & & \\
        & & & 1 & \\
        & & & & 0
    \end{bmatrix},
\end{equation}
the flux vectors are
\begin{equation}
    \mathbf{F} = \mathbf{F}_{inv} - \mathbf{F}_{vis}
\end{equation}
where the inviscid flux is
\begin{equation}
    \mathbf{F}_{inv}
    =
    \left(
    \begin{bmatrix}
        \rho u \\ \rho u^2 + p \\ \rho uv \\ \rho uw \\ u
    \end{bmatrix},
    \begin{bmatrix}
        \rho v \\ \rho uv \\ \rho v^2 + p \\ \rho vw \\ v
    \end{bmatrix},
    \begin{bmatrix}
        \rho w \\ \rho uw \\ \rho vw \\ \rho w^2 + p \\ w
    \end{bmatrix}
    \right)
\end{equation}
and the viscous flux is
\begin{equation}
    \mathbf{F}_{vis}
    =
    \left(
    \begin{bmatrix}
        0 \\
        2 \mu \frac{\partial u}{\partial x} \\
        \mu \left( \frac{\partial u}{\partial y} + \frac{\partial v}{\partial x} \right) \\
        \mu \left( \frac{\partial u}{\partial z} + \frac{\partial w}{\partial x} \right) \\
        0
    \end{bmatrix},
    \begin{bmatrix}
        0 \\
        \mu \left( \frac{\partial u}{\partial y} + \frac{\partial v}{\partial x} \right) \\
        2 \mu \frac{\partial v}{\partial y} \\
        \mu \left( \frac{\partial v}{\partial z} + \frac{\partial w}{\partial y} \right) \\
        0
    \end{bmatrix},
    \begin{bmatrix}
        0 \\
        \mu \left( \frac{\partial u}{\partial z} + \frac{\partial w}{\partial x} \right) \\
        \mu \left( \frac{\partial v}{\partial z} + \frac{\partial w}{\partial y} \right) \\
        2 \mu \frac{\partial w}{\partial z} \\
        0
    \end{bmatrix}
    \right),
\end{equation}
and the body force source term is
\begin{equation}
    \mathbf{B}
    =
    \begin{bmatrix}
        0 \\ \rho \mathbf{g}_x \\ \rho \mathbf{g}_y \\ \rho \mathbf{g}_z \\ 0
    \end{bmatrix}.
\end{equation}

\subsubsection{Eigenstructure}
The eigenstructure of the governing equations, that is, the eigenvalues and the right and left eigenvectors of the Jacobian, is important for developing a Godunov-type scheme.
Without loss of generality, consider the $x$-direction inviscid fluxes $\mathbf{F}_{x,inv}$.
The Jacobian is 
\begin{equation}
    \mathbf{A}_x = \frac{\partial \mathbf{F}_{x,inv}}{\partial \mathbf{Q}} =
    \begin{bmatrix}
        0 & 1 & 0 & 0 & 0 \\
        -u^2 & 2u & 0 & 0 & \beta \\
        -uv & v & u & 0 & 0 \\
        -uw & w & 0 & u & 0 \\
        -u/\rho & 1/\rho & 0 & 0 & 0
    \end{bmatrix}
\end{equation}
with eigenvalues
\begin{align}
    \lambda_1 = \lambda_2 = \lambda_3 = u \\
    \lambda_{4,5} = \frac12(u \pm c) 
\end{align}
where $c = \sqrt{u^2 + 4\beta/\rho}$.
Note that this means $u = \lambda_4 + \lambda_5$ and $c = \lambda_4 - \lambda_5$.
So the right and left eigenvector matrices are
\begin{gather}
    \mathbf{R}_x =
    \begin{bmatrix}
        1 & 0 & 0 & \lambda_4 & \lambda_5 \\
        u & 0 & 0 & u\lambda_4 + \beta/\rho & u \lambda_5 + \beta/\rho \\
        0 & 1 & 0 & v\lambda_4 & v\lambda_5 \\
        0 & 0 & 1 & w\lambda_4 & w\lambda_5 \\
        0 & 0 & 0 & -\lambda_5/\rho & -\lambda_4 / \rho
    \end{bmatrix}
    \\
    \mathbf{L}_x =
    \begin{bmatrix}
        1 + u^2\rho/\beta & -\rho u / \beta & 0 & 0 & -\rho \\
        u^2v\rho/\beta & -uv\rho/\beta & 1 & 0 & -\rho v \\
        u^2w\rho/\beta & -uw\rho/\beta & 0 & 1 & -\rho w \\
        -u\lambda_4\rho/\beta c & \lambda_4 \rho / \beta c & 0 & 0 & \rho/c \\
        u\lambda_5 \rho / \beta c & -\lambda_5 \rho / \beta c & 0 & 0 & -\rho/c
    \end{bmatrix}.
\end{gather}
It is also useful to define the matrix
\begin{equation}
    |\mathbf{\Lambda}_x|
    =
    \begin{bmatrix}
    |\lambda_1| & & & & \\
    & |\lambda_2| & & & \\
    & & |\lambda_3| & & \\
    & & & |\lambda_4| & \\
    & & & & |\lambda_5|
    \end{bmatrix}.
\end{equation}

\subsection{Discretisation}
\subsubsection{Finite volume method}
For transient equations such as \eqref{eq:compact_form}, artificial compressibility uses dual time stepping, that is, driving the solution to the incompressible limit each real-time step.
To advance the variables in pseudo-time, we follow the discretisation method of Vermiere et al.\ \cite{Vermeire2019}, applying it to variable-density flows for the first time.
Moving the real-time derivative to the right-hand side of the compact form \eqref{eq:compact_form}, along with the fluxes and gravity source term, yields
\begin{equation}
    \frac{\partial \mathbf{Q}}{\partial \tau}
    =
    - \nabla \cdot \mathbf{F}
    - \mathbf{I}_C \frac{\partial \mathbf{Q}}{\partial t}
    + \mathbf{B}.
\end{equation}
Integrating over the computational cell $\Omega$ and applying Gauss's theorem results in the semi-discretised equation
\begin{equation}
    \frac{\partial \mathbf{Q}_i}{\partial \tau}
    =
    - \frac{1}{|\Omega|} \sum_{f \in \partial \Omega} \mathbf{F}_f \cdot \mathbf{S}_f
    - \mathbf{I}_C \frac{\partial \mathbf{Q}_i}{\partial t}
    + \mathbf{B}_i
\end{equation}
where the subscript $i$ denotes cell averages and the subscript $f$ face averages.
Integrating over the pseudo-time interval $[\tau_0, \tau_1]$ and applying the Fundamental Theorem of Calculus, we get
\begin{equation} \label{eq:AC_rearranged}
    \mathbf{Q}_i^{m+1}
    =
    \mathbf{Q}_i^m
    +
    \int_{\tau_0}^{\tau_1} \left(
    - \frac{1}{|\Omega|} \sum_{f \in \partial \Omega} \mathbf{F}_f \cdot \mathbf{S}_f
    - \mathbf{I}_C \frac{\partial \mathbf{Q}_i}{\partial t}
    + \mathbf{B}_i \right) \, d\tau.
\end{equation}
This equation, still exact with no approximations, shows that the solution can march forward in pseudo-time with the real-time term $\mathbf{I}_C \cdot \partial \mathbf{Q}_i / \partial t$ treated as a source term.

\subsubsection{Explicit Runge-Kutta pseudo-time stepping}
The exact equation \eqref{eq:AC_rearranged} is discretised in pseudo-time using Jameson's \cite{Jameson1991} explicit Runge-Kutta multistage method, again as in Vermiere et al.\ \cite{Vermeire2019}.
Computationally, this means that there are three nested loops, as shown in Algorithm \ref{alg:pseudo_time}. 
The outer loop loops over real-time, the middle loops over pseudo-time, and the inner loops over Runge-Kutta stages.
Each real-time iteration involves enough pseudo-time iterations to get the solution to converge within a prescribed tolerance, and each pseudo-time iteration  involves a set number of Runge-Kutta stages.
The real-time derivative is updated every pseudo-time step, which takes care of the real-time stepping point-implicitly, while pseudo-time is dealt with explicitly.

\begin{algorithm}
    \caption{Three nested loops with the same array $\mathbf{Q} = [\rho, \rho u, \rho v, \rho w, p/\beta]$ operated on throughout. Backward-Euler real-time derivative shown for illustration; higher-order methods require more real-time steps saved than $\mathbf{Q}^n$.}\label{alg:pseudo_time}
    \begin{algorithmic}
    \State $t \gets 0$
    \While{$t<t_{max}$} \Comment{start real-time loop}
        \State $\mathbf{Q}^n \gets \mathbf{Q}$
        \State $t \gets t + \Delta t$
        \State $\tau \gets 0$
        
        \While{residuals $>$ tolerance} \Comment{start pseudo-time loop}
            \State $\mathbf{Q}^m \gets \mathbf{Q}$
            \State $\tau \gets \tau + \Delta \tau$
            \State real-time term $\gets \mathbf{I}_C \frac{\mathbf{Q} - \mathbf{Q}^n}{\Delta t}$
            \State body force $\gets \rho \mathbf{g}$
            
            \For{$s = 1, ..., s_{max}$} \Comment{start Runge-Kutta loop}
                \State calculate $\alpha_{s}, \alpha_{PI}, \mathbf{F}_f$
                \State $\mathbf{Q} \gets \mathbf{Q}^m
                + \Delta \tau \frac{\alpha_s}{\alpha_{PI}}\left(
                - \frac{1}{|\Omega|} \sum_f \mathbf{F}_f \cdot \mathbf{S}_f 
                - \text{real-time term} + \text{body force}
                \right)$

            \EndFor \Comment{end Runge-Kutta loop}
        \EndWhile \Comment{end pseudo-time loop}
    \EndWhile \Comment{end real-time loop}
    \end{algorithmic}
\end{algorithm}

Formally, let $\mathbf{Q}^{n,m,s}$ denote the conserved variables $\mathbf{Q}$ evaluated at the $n$th real-time step, the $m$th pseudo-time iteration and the $s$th stage in the Runge-Kutta scheme.
The variables are advanced in pseudo-time using the update equation
\begin{align}
    \label{eq:RK1}
    \mathbf{Q}_i^{n+1,m+1,0} &= \mathbf{Q}_i^{n+1,m}
    \\
    \label{eq:RK2}
    \mathbf{Q}_i^{n+1,m+1,s} &=
    \begin{multlined}[t]
        \mathbf{Q}_i^{n+1,m+1,0}
        \\ +
        \frac{\alpha_s \Delta \tau}{\alpha_{PI}}
        \left(
        - \frac{1}{|\Omega|} \sum_{f \in \partial \Omega} \mathbf{F}_f^{n+1,m+1,s-1} \cdot \mathbf{S}_f
        - \mathbf{I}_C \frac{\partial}{\partial t} (\mathbf{Q}_i^{n+1,m+1,0})
        + \mathbf{B}_i^{n+1,m+1,0}
        \right)
        \\ \text{for } s = 1,...,s_{max}
    \end{multlined}
    \\
    \mathbf{Q}_i^{n+1,m+1} &= \mathbf{Q}_i^{n+1,m+1,s_{max}}
\end{align}
where $\alpha_s$ is the Runge-Kutta coefficient.
The last coefficient $\alpha_{s_{max}}$ must be equal to $1$ and, for second-order accuracy in time, the second-last coefficient $\alpha_{s_{max}-1}$ must be equal to $1/2$.

The real-time derivative in \eqref{eq:RK2} is given by the backward differencing scheme
\begin{equation} \label{eq:bdf}
    \frac{\partial}{\partial t} \left(\mathbf{Q}_i^{n+1,m+1,0}\right)
    = \frac{B_0 \mathbf{Q}_i^{n+1,m+1,0} + \sum_{\sigma = 0}^M B_{\sigma + 1} \mathbf{Q}_i^{n-\sigma}}{\Delta t}
\end{equation}
where the coefficients $B_{\sigma}$ are given in Table \ref{tab:BDF}.
Note that the first real-time derivative can only be Backward-Euler, the second real-time derivative Backward-Euler or BDF2, and so on.
In any case, we only use Backward-Euler in this paper.
The effect of having the first term on the right-hand side of \eqref{eq:bdf} be $\mathbf{Q}_i^{n+1,m+1,0}$ and not $\mathbf{Q}_i^{n+1,m+1,s}$ is remedied exactly with the ``point-implicit scaling coefficient''
\begin{equation} \label{eq:point_implicit}
    \alpha_{PI} = 1 + B_0 \alpha_{S} \frac{\Delta \tau}{\Delta t}
\end{equation}
which comes from some rearranging and helps with stability when $\Delta \tau/ \Delta t$ is large \cite{Loppi2019, Vermeire2019}.
Vermiere et al.\ \cite{Vermeire2019} set $\alpha_{PI} = 1$ for constant density flows but, in the present study of variable density flows, it was vital for stability to use the correct value \eqref{eq:point_implicit}.

\begin{table}[ht]
    \centering
    \caption{Backward differencing coefficients from Vermiere et al.\ \cite[Table 1]{Vermeire2019} with typo fixed according to the standard textbook \cite[p.27]{Iserles2008}.}
    \begin{tabular}{
     l c c c c 
    }
         \hline
         & $B_0$ & $B_1$ & $B_2$ & $B_3$ \\
         \hline
         Backward-Euler & $1$ & $-1$ & $-$ & $-$ \\
         BDF2 & $\frac32$ & $-2$ & $\frac12$ & $-$ \\
         BDF3 & $\frac{11}{6}$ & $-3$ & $\frac32$ & $-\frac13$ \\
         \hline
    \end{tabular}
    \label{tab:BDF}
\end{table}

An explicit scheme was chosen for pseudo-time because explicit schemes have numerical advantages compared to implicit schemes.
Indeed, while implicit schemes are unconditionally stable, they are less accurate than explicit schemes when large time steps (or Courant numbers) are used \cite{Fernandez2018}.
Moreover, in implicit schemes it is impossible to determine in advance the upper limit of the Courant number before the accuracy starts to decrease \cite{Morales2014}.

The specific time discretisation chosen also has computational advantages.
First, as the scheme is explicit in pseudo-time, it is easy to parallelise \cite{Loppi2019,Vermeire2019}.
Second, as it is point-implicit in real-time, large real-time steps are possible \cite{Loppi2019}.
Third, the specific multistage Runge-Kutta method it uses has low memory requirements.
Essentially, the same array $\mathbf{Q}^{n,m,s}$ gets operated on through the real-time, pseudo-time, and Runge-Kutta loops.
Only the initial Runge-Kutta stage \eqref{eq:RK1} and the bracketed part of the update equation \eqref{eq:RK2} for the latest iteration need to be stored \citep{Jameson1991,Blazek2005}.
Consequently, there is potential for the method to be implemented very efficiently.

\subsubsection{Local pseudo-time steps}
As the pseudo-time stepping is explicit, there is a stability restriction of
\begin{equation}
    \text{CFL}_{\tau} \leq
    \begin{cases}
    1 &\text{ if 1D}\\
    \frac12 &\text{ if 2D}\\
    \frac14 &\text{ if 3D}
    \end{cases}
\end{equation}
where $\text{CFL}_{\tau}$ is the Courant number for pseudo-time.
However, slightly smaller Courant numbers (smaller by around $0.1$--$0.5$) may be required to account for the fact that the restriction is based on the last pseudo-time step and velocities may increase in the current pseudo-time step.

Using one global time step constrained by the most extreme restriction across the mesh will slow down convergence substantially.
To see this, consider a variable-density case with a high density ratio, such as water and air.
The acoustic wave speed will be much lower through the water than the air, and consequently pseudo-time convergence will be much slower in the water than the air \cite{Pelanti2017}.
Local pseudo-time stepping allows the water and air to converge at similar speeds while still meeting the stability restriction in each cell.

In the absence of viscosity, a stable local pseudo-time step can be given by
\begin{equation}
    \Delta \tau_i
    = \text{CFL}_{\tau} \frac{|\Omega_i|}{(\hat{\Lambda}_{\tau,x} + \hat{\Lambda}_{\tau,y} + \hat{\Lambda}_{\tau,z})_i}
\end{equation}
where
\begin{align} \label{eq:CFLcon1a}
    \hat{\Lambda}_{\tau,x} &= \frac12 (|u| + c) \Delta \hat{S}_x \\ \label{eq:CFLcon1b}
    \hat{\Lambda}_{\tau,y} &= \frac12 (|v| + c) \Delta \hat{S}_y\\ \label{eq:CFLcon1c}
    \hat{\Lambda}_{\tau,z} &= \frac12 (|w| + c) \Delta \hat{S}_z
\end{align}
and
\begin{align} \label{eq:CFLcon2a}
    \Delta \hat{S}_x &= \frac12 \sum_{f} |S_x|_f \\ \label{eq:CFLcon2b}
    \Delta \hat{S}_y &= \frac12 \sum_{f} |S_y|_f \\ \label{eq:CFLcon2c}
    \Delta \hat{S}_z &= \frac12 \sum_{f} |S_z|_f 
\end{align}
define the spectral radii for the convective fluxes \citep{Blazek2005}.
The division by two in \eqref{eq:CFLcon1a}--\eqref{eq:CFLcon1c} is because of the definition of $\lambda_{4,5}$, while the division by two in \eqref{eq:CFLcon2a}--\eqref{eq:CFLcon2c} is because $\Delta \hat{S}$ is the projection of the entire computational cell $\Omega$ onto the relevant plane.

\subsubsection{Global real-time steps}
As real-time is treated point-implicitly, there is no stability restriction as for pseudo-time.
Nonetheless, it is useful to be able to set how relatively large the real-time steps are.
So, the global real-time step can be given by
\begin{equation}
    \Delta t
    = \min_i \left (\text{CFL}_t \frac{|\Omega_i|}{(\hat{\Lambda}_{t,x} + \hat{\Lambda}_{t,y} + \hat{\Lambda}_{t,z})_i} \right)
\end{equation}
where $\text{CFL}_t$ is the Courant number for real-time,
\begin{align}
    \hat{\Lambda}_{t,x} &= |u| \cdot \Delta \hat{S}_x \\
    \hat{\Lambda}_{t,y} &= |v| \cdot \Delta \hat{S}_y \\
    \hat{\Lambda}_{t,z} &= |w| \cdot \Delta \hat{S}_z
\end{align}
and $\Delta \hat{S}_x, \Delta \hat{S}_y, \Delta \hat{S}_z$ are as previously defined in \eqref{eq:CFLcon2a}--\eqref{eq:CFLcon2c}.

\subsection{Reconstruction at cell interfaces}
In this study, we use standard MUSCL reconstruction to find Riemann states of density and velocity at each cell interface, as in Kelecy and Pletcher \cite{Kelecy1997} and Qian et al.\ \cite{Qian2006}.
However, special treatment is required to reconstruct pressure at the free surface as the density discontinuity induces a pressure gradient discontinuity.
This section outlines the standard and pressure-specific approaches in turn.

\subsubsection{Standard method} \label{sec:standard}
The primitive variables $\mathbf{W}$ are reconstructed at the interface $f$ between the cells $i$ and $j$ to give the left and right values
\begin{gather} \label{eq:muscl1}
    \mathbf{W}_L = \mathbf{W}_i + \nabla \mathbf{W}_i \cdot (\mathbf{W}_f - \mathbf{W}_i)\\ \label{eq:muscl2}
    \mathbf{W}_R = \mathbf{W}_j + \nabla \mathbf{W}_j \cdot (\mathbf{W}_f - \mathbf{W}_j)
\end{gather}
where a TVD limiter must be applied to $\nabla \mathbf{W}$ to avoid Gibbs-type oscillations.
In the $x$-direction of a structured mesh, this reduces to
\begin{equation}
    \nabla \mathbf{W}_i
    = G \left( 
    \frac{\mathbf{W}_{i+1} - \mathbf{W}_i}{dx/2},
    \frac{\mathbf{W}_{i} - \mathbf{W}_{i-1}}{dx/2}
    \right)
\end{equation}
where $G$ is the limiter function.
Unstructured meshes need a slightly different approach \cite{Barth1989,Venkatakrishnan1995}, but that will not be considered further here.
In this study, we use the $k$-limiter
\begin{gather}
    G(a,b) = s \cdot \max(0, \min(k|b|, s\cdot a), \min(|b|, ks\cdot a)) \\ s = \sgn(b)
\end{gather}
where $k=1$ is the minmod limiter and $k=2$ the superbee limiter \cite{Qian2006}.
This standard limiter is suitable for reconstructing the density and velocity.

\subsubsection{Pressure gradient} \label{sec:gradp}
Using a standard limiter to reconstruct pressure can result in an imbalance of forces at the free surface and, in turn, either parasitic velocities \cite{DeGroot2011,Efremov2017,Hashemi2020} or compression at what should be the incompressible limit.
Only Qian et al.\ \cite{Qian2006} have addressed this problem for the variable-density artificial-compressibility equations, and only in the hydrostatic case.
However, the problem persists in the absence of gravity.
Both the hydrostatic and non-hydrostatic problems are considered below, before outlining a new method for calculating the pressure gradient based on a rearrangement of the momentum equation.

\paragraph{Hydrostatic}
First, consider a stationary column of air overlaying a stationary column of water under the influence of gravity.
The momentum equation should settle to $\nabla p = \rho \mathbf{g}$ at the incompressible limit, that is, a hydrostatic pressure distribution.
A scheme that can reproduce such a steady state (up to machine precision) is called well-balanced \cite{Botta2004}.

It is the free surface that complicates matters.
The jump in density leads to a jump in $\nabla p$ which, if not calculated carefully, can lead to spurious or parasitic velocities \cite{DeGroot2011,Efremov2017,Hashemi2020}.
In a Godunov-type scheme, the problem arises because the gradients calculated in the MUSCL reconstruction step indiscriminately include information from both sides of the free surface.
No matter the order of interpolation or refinement of the grid, the MUSCL reconstruction phase
will always have this problem, but the impact will be on a different scale \cite{Krause2019}.
Broadly, there are two ways to deal with this: either at the global level or the local level
\cite{Zingale2002}.
The global method removes hydrostatic pressure from the governing equations, while the local method involves including hydrostatic pressure in the reconstruction phase \cite{Qian2006,Botta2004,Krause2019,Fuchs2010}.

In their local method, Qian et al.\ \cite{Qian2006} split the full (dynamic) pressure gradient into hydrostatic and kinematic parts
\begin{equation}
    \nabla p^{dyn} = \nabla p^{hyd} + \nabla p^{kin}.
\end{equation}
For structured meshes, they first calculated the dynamic gradients $\nabla p^{dyn}_{i,L}$ and $\nabla p^{dyn}_{i,R}$ with one-sided differences.
They then subtracted the hydrostatic gradients $\nabla p^{hyd}_{i,L} = (\rho_{i-1} + \rho_i) \mathbf{g}/2$ and $\nabla p^{hyd}_{i,R} = (\rho_i + \rho_{i+1}) \mathbf{g}/2$ from the dynamic gradients $\nabla p^{dyn}_{i,L}$ and $\nabla p^{dyn}_{i,R}$ to get the kinematic gradients $\nabla p^{kin}_{i,L}$ and $\nabla p^{kin}_{i,R}$.
They then applied a standard limiter to the kinematic gradients $\nabla p^{kin}_{i,L}$ and $\nabla p^{kin}_{i,R}$ to get the limited kinematic gradient $\nabla p^{kin}_i$.
Finally, they added the hydrostatic gradient $\nabla p^{hyd}_i = \rho_i \mathbf{g}$ to the limited kinematic gradient $\nabla p^{kin}_i$ to get the limited dynamic pressure gradient $\nabla p^{dyn}_i$.
This method successfully resolves the pressure gradient discontinuity in the hydrostatic case, keeping the scheme well-balanced.
However, as it is complicated to implement standard limiters on unstructured meshes \cite{Barth1989,Venkatakrishnan1995}, it is even more complicated for this limiter.
A more easily implementable limiter would be preferable for practical applications that require unstructured grids.
Moreover, this limiter does not address non-hydrostatic pressure gradient discontinuities.
Nonetheless, it was implemented in this paper for the purpose of comparison.

\paragraph{Non-hydrostatic}
There is still a pressure gradient discontinuity in the absence of gravity.
Let us denote the jump in any variable $x$ over the free surface by $[x]$.
The kinematic boundary condition, given by $[\mathbf{u}] = 0$, can be combined with the momentum equation to derive the dynamic boundary condition, given by $[\nabla p/\rho] = 0$  \citep{Ntouras2020,Queutey2007,Kruisbrink2018,Vukcevic2018}.
Since $[\rho] \neq 0$, we must also have $[\nabla p] \neq 0$ to satisfy the dynamic boundary condition.
Kruisbrink et al.\ \cite{Kruisbrink2018} bypassed this problem by counteracting spurious forces with a quasi-buoyancy force, while Vukčević et al.\ \cite{Vukcevic2018} used the ghost fluid method to interpolate variables in a one-sided manner consistent with the dynamic boundary condition.
The difficulty with these methods is that they require a method to detect the free surface, which is ambiguous in pseudo-time as the water is allowed to artificially compress.
Alternatively, the pressure gradient can be normalised by the density \cite{Ntouras2020,Queutey2007}.

\paragraph{New method}
Here, we introduce a new method for calculating the pressure gradient.
Without loss of generality, consider the $x$-direction momentum equation in the absence of viscosity.
Note that the pseudo-time derivatives disappear at the incompressible limit.
The momentum equation can then be rearranged, giving
\begin{equation} \label{eq:new_gradp}
    \frac{\partial p}{\partial x}
  = -\frac{\partial}{\partial t}(\rho u)
    -\frac{\partial}{\partial x}(\rho u^2)
    -\frac{\partial}{\partial y}(\rho u v)
    -\frac{\partial}{\partial z}(\rho u w)
    +\rho\mathbf{g}_x.
\end{equation}
All the terms on the right-hand side are known.
First, the real-time derivative $\partial(\rho u)/\partial t$ is already calculated as a source term.
Second, the spatial derivatives can be expanded using the product and chain rules to get
\begin{gather}
    \frac{\partial}{\partial x} (\rho u^2) = u^2 \frac{\partial}{\partial x} (\rho) + 2 \rho u \frac{\partial}{\partial x} (u)
     \\
    \frac{\partial}{\partial y} (\rho uv) = uv \frac{\partial}{\partial y} (\rho) + \rho u \frac{\partial}{\partial y} (v) + \rho v \frac{\partial}{\partial y} (u)
     \\
    \frac{\partial}{\partial z} (\rho uw) = uw \frac{\partial}{\partial z} (\rho) + \rho u \frac{\partial}{\partial z} (w) + \rho w \frac{\partial}{\partial z} (u)
\end{gather}
where the density and velocity are set explicitly, and the derivatives $\partial / \partial x$, $\partial / \partial y$, and $\partial / \partial z$ are given by the TVD gradients already calculated for the density and velocity.

This pressure gradient calculation takes advantage of the real-time source term to ensure that the TVD interpolation satisfies the differential form of the momentum equation at the incompressible limit.
Moreover, when the velocity is zero, the momentum equation \eqref{eq:new_gradp} reduces to
\begin{equation}
    \frac{\partial p}{\partial x} = \rho\mathbf{g}_x,
\end{equation}
that is, local hydrostatic reconstruction, and so the method is well-balanced.
Furthermore, it is straightforward to implement on unstructured meshes, unlike the split limiter.
Of course, it does not include the pseudo-time term as pseudo-time is explicit.
However, as long as there is convergence over pseudo-time, it is the incompressible limit that matters rather than how the solution is reached.

\subsection{Flux calculation}
The inviscid flux $\mathbf{F}_{inv}$ is calculated with a Riemann solver.
This study compares the Bassi, Roe, and Osher Riemann solvers.
The viscous flux $\mathbf{F}_{vis}$ can be calculated easily using standard finite difference methods \cite{Bhat2019,Kelecy1997}, but the effect of viscosity is ignored in this paper as only convection-dominated flows are considered in the benchmark tests.

First note that, for cell faces with normal vectors pointing in any other direction than the $x$-axis, we use 3D rotation matrices to convert the Riemann problem to a $x$-direction Riemann problem as in \cite{Tanaka1994}.
At each face, the rotation matrix is given by
\begin{equation}
    \mathbf{R} =
    \begin{bmatrix}
    1 & & & & \\
    & n_x & n_y & n_z & \\
    & t_{1,x} & t_{1,y} & t_{1,z} & \\
    & t_{2,x} & t_{2,y} & t_{2,z} & \\
    & & & & 1
    \end{bmatrix}
\end{equation}
where $\mathbf{n}$ is the unit normal face vector, and $\mathbf{t}_1$ and $\mathbf{t}_2$ are unit vectors tangential to $\mathbf{n}$, making an orthonormal set $\{\mathbf{n}, \mathbf{t}_1, \mathbf{t}_2\}$.
Given any face vector $\mathbf{n}$, there are infinite possibilities for $\mathbf{t}_1$ and $\mathbf{t}_2$---one possibility is given in Algorithm \ref{alg:tangential}.
The Riemann states are then $\mathbf{Q}_L = \mathbf{R} \cdot \mathbf{Q}_a$ and $\mathbf{Q}_R = \mathbf{R} \cdot \mathbf{Q}_b$, assuming $\mathbf{n}$ points out of cell $a$ and into cell $b$.
After flux calculation, we use the inverse rotation matrices to transform the flux back to the original reference frame.
As $\mathbf{R}$ is a rotation matrix, its inverse is $\mathbf{R}^{-1} = \mathbf{R}^T$.
Thus the flux is
\begin{equation}
    \mathbf{F}_{inv} = \mathbf{R}^T \cdot \mathbf{F}_{x,inv}(\mathbf{R} \cdot \mathbf{Q}_a,\mathbf{R} \cdot \mathbf{Q}_b)
\end{equation}
where $\mathbf{F}_{x,inv}(\mathbf{Q}_L, \mathbf{Q}_R)$ is the Riemann solver for the $x$-direction inviscid fluxes.

\begin{algorithm}
    \caption{One possibility for calculating $\mathbf{t}_1$ and $\mathbf{t}_2$ from $\mathbf{n}$.}\label{alg:tangential}
    \begin{algorithmic}
    \If{$n_x == 0$}
        \State $\mathbf{t}_1 \gets (1,0,0)$
    \ElsIf{$n_y == 0$}
        \State $\mathbf{t}_1 \gets (0,1,0)$
    \Else
        \State $\mathbf{t}_1 \gets \frac{1}{\sqrt{n_x^2 + n_y^2}}(n_y,-n_x,0)$
    \EndIf
    \State $\mathbf{t}_2 \gets \mathbf{n} \times \mathbf{t}_1$
    \end{algorithmic}
\end{algorithm}

\subsubsection{Bassi Riemann solver} \label{sec:bassi}
Bassi et al.\ \cite{Bassi2018} assumed that the contact wave lies between the acoustic (shock and/or rarefaction) waves to derive the Riemann invariants and Rankine-Hugoniot conditions for the variable-density artificial-compressibility equations.
They found that the equations have the unusual property of shocks and rarefactions travelling at the same speed, with the head and tail of the rarefaction coinciding, and the values of the conserved variables in the star region independent of the wave type.
That is, shocks and rarefactions are equivalent, which means that the exact solution of the Riemann problem can be found non-iteratively.
Consequently, using such an algorithm to calculate the numerical flux at cell interfaces (i.e.\ using an exact Riemann solver) is not as prohibitively expensive as for other governing equations.

The exact Riemann solver is as follows.
Without loss of generality, for the $x$-split equations, the left and right acoustic wave speeds are
\begin{gather}
    \label{eq:bassi_start}
    S_L = \frac12 (u_L - c_L) \\
    S_R = \frac12 (u_R + c_R)
\end{gather}
regardless of whether they are shocks or rarefactions.
In the star region, we have
\begin{gather}
    u_* = \frac{p_R - p_L + \rho_R u_R \lambda_{5,R} - \rho_L u_L \lambda_{4,L}}{\rho_R \lambda_{5,R} - \rho_L \lambda_{4,L}} \\
    p_* = p_R + \rho_R \lambda_{5,R} (u_R - u_*) \\
    \rho_{*L} = \frac{\rho_L \lambda_{4,L}}{u_* - \lambda_{5,L}} \\
    \rho_{*R} = \frac{\rho_R \lambda_{5,R}}{u_* - \lambda_{4,R}}
    \label{eq:bassi_end}
\end{gather}
again regardless of whether the acoustic waves are shocks or rarefactions, and the tangential velocities only change across the contact wave travelling with speed $u_*$, that is,
\begin{gather}
    v_{*L} = v_L \\
    v_{*R} = v_R \\
    w_{*L} = w_L \\
    w_{*R} = w_R.
\end{gather}
The exact Riemann solver determines the location of the waves with respect to the $\tau$-axis, and then substitutes the appropriate values of density, velocity, and pressure into the flux function $\mathbf{F}_{x,inv}$.

As this exact solution is so simple, it has been used to calculate numerical fluxes in some discontinuous Galerkin models \cite{Bassi2018,Manzanero2020}.
However, Bassi et al.\ \cite[Appendix B4]{Bassi2018} did acknowledge that it is possible for the contact wave to overtake the acoustic wave, meaning that the assumptions break down and the Riemann solver is no longer exact.
Consequently, approximate Riemann solvers that do not make this assumption must be investigated as well.

\subsubsection{Roe Riemann solver} \label{sec:roe}
The Roe Riemann solver \cite{Roe1981} is an approximate method that captures the contact wave.
It is given by
\begin{equation}
    \mathbf{F}_{x,inv} = \frac12 \left(\mathbf{F}_{x,inv}(\mathbf{Q}_R) + \mathbf{F}_{x,inv}(\mathbf{Q}_L) - |\mathbf{\tilde{A}}_x|\left(\mathbf{Q}_R - \mathbf{Q}_L\right)\right)
\end{equation}
where
\begin{equation}
    |\mathbf{\tilde{A}}_x| = \mathbf{\tilde{R}}_x |\mathbf{\tilde{\Lambda}}_x| \mathbf{\tilde{L}}_x
\end{equation}
is evaluated at $\mathbf{\tilde{Q}}$ and the tilde denotes the Roe average.
Kelecy and Pletcher \cite{Kelecy1997} and Qian et al.\ \cite{Qian2006} used this Riemann solver for the variable-density artificial-compressibility equations, with the Roe averages
\begin{gather}
    \tilde{\rho} = \sqrt{\rho_L \rho_R} \\
    \tilde{u} = \frac{\sqrt{\rho_L} u_L + \sqrt{\rho_R} u_R}{\sqrt{\rho_L} + \sqrt{\rho_R}} \\
    \tilde{v} = \frac{\sqrt{\rho_L} v_L + \sqrt{\rho_R} v_R}{\sqrt{\rho_L} + \sqrt{\rho_R}} \\
    \tilde{w} = \frac{\sqrt{\rho_L} w_L + \sqrt{\rho_R} w_R}{\sqrt{\rho_L} + \sqrt{\rho_R}}.
\end{gather}
Pressure does not appear in the eigenstructure so it does not need a Roe average.
This Riemann solver has also been used for the artificial compressibility equations paired with VOF \cite{Pan2000,Ntouras2020}.

Generally, the Roe Riemann solver should have an entropy fix to account for transonic rarefactions \cite{Toro2009}, but this was not considered by either Kelecy and Pletcher \cite{Kelecy1997} or Qian et al.\ \cite{Qian2006}.
However, as mentioned above, the rarefaction fans for the variable-density artificial-compressibility equations have the strange property of the head and tail speeds coinciding \cite{Bassi2018}.
Consequently, transonic rarefactions are impossible, and so an entropy fix is not needed.

\subsubsection{Osher Riemann solver} \label{sec:osher}
The Osher Riemann solver \cite{Osher1982} is another approximate method that captures the contact wave.
It is given by
\begin{equation}
    \mathbf{F}_{x,inv} = \frac12 \left(\mathbf{F}_{x,inv}(\mathbf{Q}_R) + \mathbf{F}_{x,inv}(\mathbf{Q}_L) - \int_{\mathbf{Q}_L}^{\mathbf{Q}_R} |\mathbf{A}_x|\, d\mathbf{Q} \right)
\end{equation}
where the integration path must be determined.
Traditionally, the idea is to construct a path passing through each of the intermediate states in the wave structure, but this can be complicated.
Dumbser and Toro \cite{Dumbser2011} provide a simpler method that only requires the eigenstructure of the equations, and not the specific wave structure of the Riemann problem.
They integrate along a straight line in phase space using three-point Gaussian quadrature
\begin{equation}
    \int_{\mathbf{Q}_L}^{\mathbf{Q}_R} |\mathbf{A}_x|\, d\mathbf{Q}
    = \frac12 \left(\mathbf{Q}_R - \mathbf{Q}_L \right) \sum_{j=1}^3 w_j |\mathbf{A}_{x,j}|
\end{equation}
where 
\begin{equation}
    |\mathbf{A_{x,j}}| = \mathbf{R_{x,j}} |\mathbf{\Lambda_{x,j}}| \mathbf{L_{x,j}}
\end{equation}
is evaluated at
\begin{equation}
    \mathbf{Q}_j := \mathbf{Q}_L + (\mathbf{Q}_R - \mathbf{Q}_L) \left( \frac12 (\xi_j + 1) \right)
\end{equation}
with the integration points $\xi_1 = -\sqrt{3/5}, \xi_2 = 0, \xi_3 = \sqrt{3/5}$ and corresponding weights $w_1 = 5/9, w_2 = 8/9, w_3 = 5/9$.
This simplification of the Osher Riemann solver has been shown to be robust for the Baer-Nunziato equations \cite{Dumbser2011}, the Euler equations \cite{Lee2013}, and the shallow water equations \cite{Glenis2018}, all of which feature a contact wave.
In this paper, we implement this Riemann solver for the variable-density artificial-compressibility equations for the first time.

\subsection{Boundary conditions} \label{sec:boundary}
The boundary conditions are implemented with ghost cells, rather than prescribing the boundary flux directly.
This means that the Riemann solver automatically calculates the correct flux.
At walls, a zero-gradient condition is applied to density and a no-slip condition to velocity, although the effect of negating the tangential velocity will have no effect in the absence of viscosity.

\begin{figure}
    \centering
    \includegraphics[width = 0.65 \textwidth]{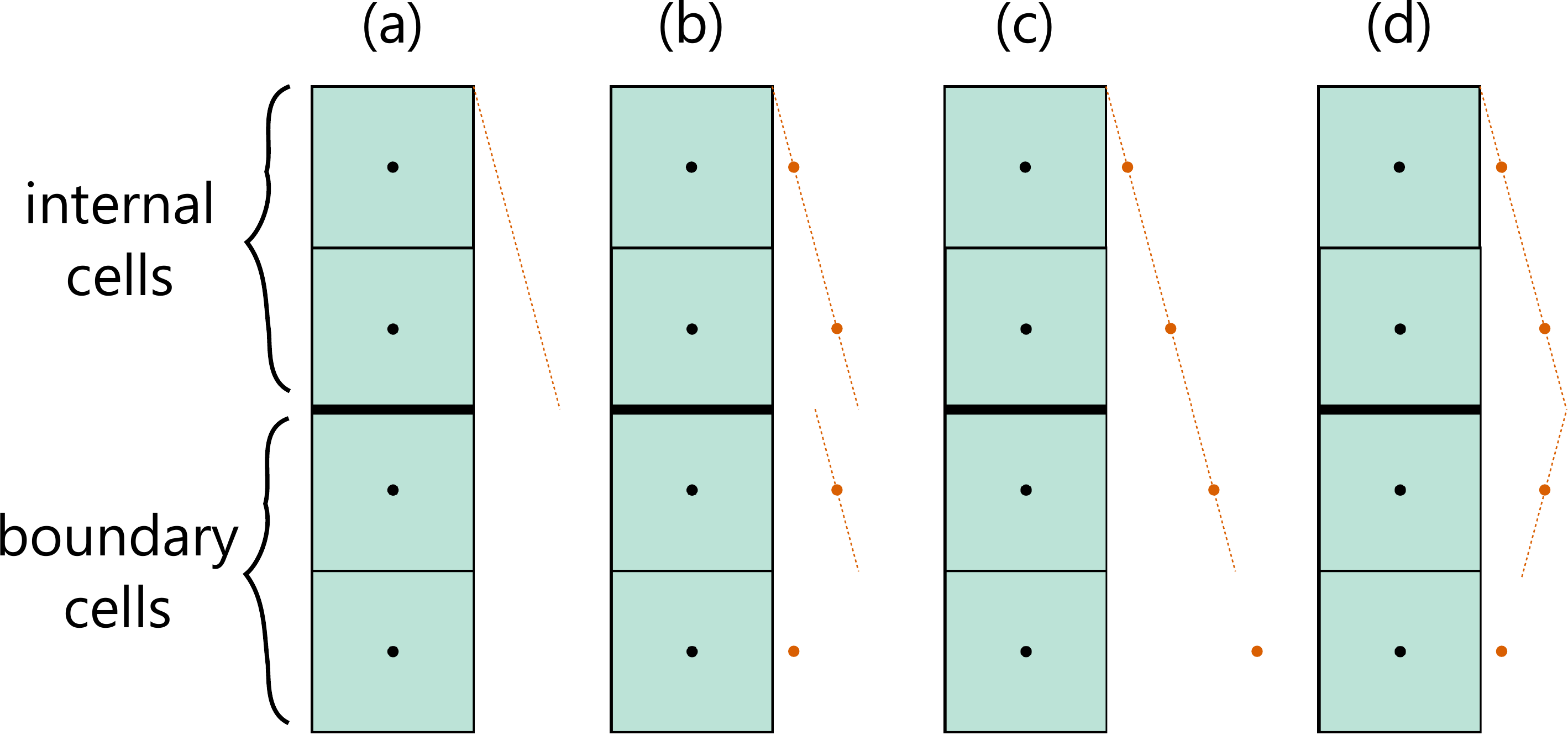}
    \caption{Options for setting pressure in the ghost cells, illustrated here for a hydrostatic case. From left to right: (a) reality, (b) zero-gradient boundary condition with local hydrostatic reconstruction, (c) linear or hydrostatic extrapolation boundary condition, (d) zero-gradient boundary condition with local hydrostatic reconstruction (reversed gravity).}
    \label{fig:pressure_ghosts}
\end{figure}

Pressure, again, is not so simple.
Consider the hydrostatic case, as in Figure \ref{fig:pressure_ghosts}.
A zero-gradient condition naively applied to pressure, as in \cite{Bhat2019}, will underestimate pressure at the lower wall, meaning that there is not enough support for the fluid above \cite{Zingale2002}, and so the solution will settle to the wrong value at the incompressible limit.
One way to overcome this is to interpolate the pressure to the ghost cells, either linearly \cite{Bhat2019a} or hydrostatically \cite{Qian2006}.
However, interpolation has disadvantages.
First, it requires calculating the gradient twice: once for the MUSCL reconstruction and once for ghost cells.
Second, for hydrostatic interpolation, while the normal velocity is indeed zero at the wall, it is not zero throughout the whole cell, and so the kinematic pressure gradient is ignored.
In this paper, we use an alternative method, acceptable only when gravity forms part of the pressure gradient calculation in the MUSCL reconstruction stage, which is indeed the case for the split pressure \cite{Qian2006} and new incompressible limit methods.
The idea is to apply a zero-gradient condition to pressure but reverse the gravity direction in the ghost cells when calculating the pressure gradient.
This automatically incorporates both the hydrostatic and kinematic pressure throughout the whole cell and means that the gradient only needs to be calculated once.

\section{Benchmark tests}
The numerical method was implemented in Python for 2D Cartesian meshes.
This implementation includes all three Riemann solvers outlined above: Bassi (Section \ref{sec:bassi}), Roe (Section \ref{sec:roe}), and Osher (Section \ref{sec:osher}).
It also includes five gradient calculations for the MUSCL reconstruction step: minmod and superbee (Section \ref{sec:standard}), hydrostatic splitting with minmod and superbee applied to the kinematic pressure (Section \ref{sec:gradp}), and the new pressure gradient method valid at the incompressible limit (Section \ref{sec:gradp}).
The following benchmark tests illustrate how different Riemann solvers and pressure gradients behave, as well as demonstrating the validity of the general method in modelling free-surface flow.
Viscosity and turbulence are ignored throughout and the real-time backward differencing scheme, if used, is set to Backward-Euler.
This allows the focus to be on the developments of this paper: Riemann solvers and pressure gradients.

\subsection{Riemann problems}
To compare how the different Riemann solvers perform, five 1D Riemann problems were used as benchmark tests, as in Bassi et al.\ \cite{Bassi2018}.
Table \ref{tab:riemann_problems} shows the initial conditions, artificial compressibility parameter $\beta$, and the output pseudo-time $\tau$ for each Riemann problem.
RP1 is Elsworth and Toro's \cite{Elsworth1992} test for the constant-density artificial-compressibility equations, designed to assess the robustness of the Riemann solvers, even though smooth velocity and pressure fields are generally expected in the artificial compressibility method.
RP2--RP4 come from Toro \cite{Toro2009};
RP2 assesses how the Riemann solvers deal with a large pressure jump, RP3 a stationary contact wave, and RP4 a slowly moving contact wave.
RP5 is RP1 but with a large density jump, devised by Bassi et al.\ \cite{Bassi2018} to test how the Riemann solvers might perform at a free surface.
Since all five Riemann problems involve a contact wave bounded by two acoustic waves, the exact solution \eqref{eq:bassi_start}--\eqref{eq:bassi_end} can be used to check the results.

\begin{table}[ht]
    \centering
    \caption{Initial conditions for Riemann problems, based on \cite[Tables 3--4]{Bassi2018}.}
    \begin{tabular}{
    c
    S[table-format=1.5]
    S[table-format=1.1]
    S[table-format=1.1]
    S[table-format=1.1]
    S[table-format=3.1]
    S[table-format=1.2]
    S[table-format=3.2]
    S[table-format=1.3]
    }
         \hline
         & $\rho_L$ & $\rho_R$ & $u_L$ & $u_R$ & $p_L$ & $p_R$ & $\beta$ & $\tau$ \\
         \hline
         1 & 1.0 & 1.0 & 1.0 & 1.0 & 0.1 & 1.0 & 0.81 & 0.1 \\
         2 & 1.0 & 1.0 & 0.0 & 0.0 & 1000.0 & 0.01 & 1000.0 & 0.005 \\
         3 & 1.4 & 1.0 & 0.0 & 0.0 & 1.0 & 1.0 & 1.0 & 2.0 \\
         4 & 1.4 & 1.0 & 0.1 & 0.1 & 1.0 & 1.0 & 1.0 & 2.0 \\
         5 & 0.00125 & 1.0 & 1.0 & 1.0 & 0.1 & 1.0 & 1.0 & 0.005 \\
         \hline
    \end{tabular}
    \label{tab:riemann_problems}
\end{table}

As the implementation is 2D, a mesh was created with 100 cells in the $x$-direction and one cell in the $y$-direction, with a zero-gradient boundary condition applied to all variables.
There is no gravity term.
For the pseudo-time stepping, which is explicit, three Runge-Kutta stages and a Courant number of $0.98$ were used with global time stepping, while the real-time term was not activated.
The minmod slope limiter was used for density, velocity, and pressure.
Minmod was used for pressure because the new method is not accurate in pseudo-time, only at the incompressible limit, and the split method is identical to the unsplit method in the absence of gravity.
The only difference between the simulations was the Riemann solver.

\begin{figure}
    \centering
    \includegraphics[width = \textwidth]{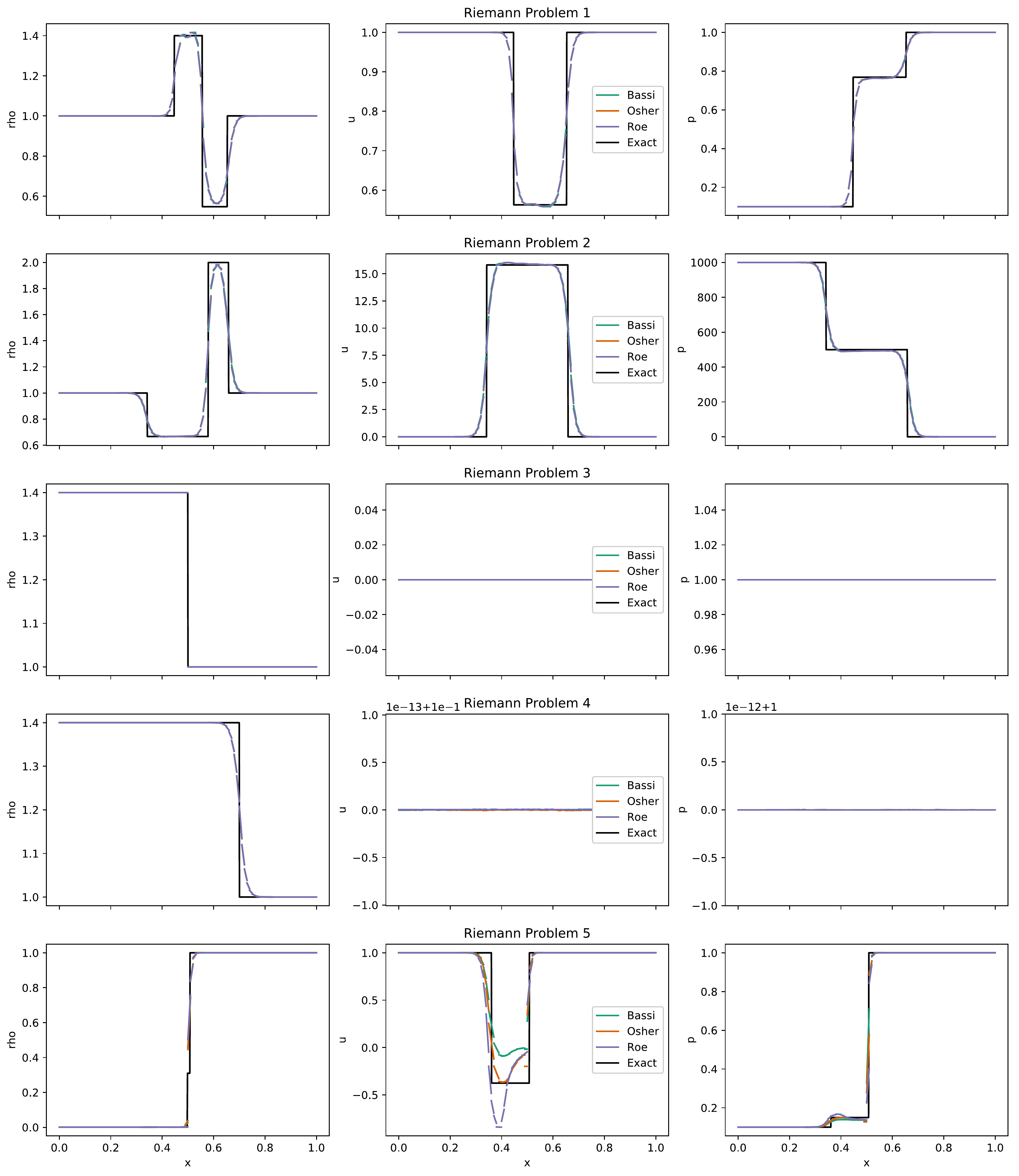}
    \caption{Comparison of Riemann solvers in 1D Riemann problems.}
    \label{fig:riemann_problems}
\end{figure}

As shown in Figure \ref{fig:riemann_problems}, all three Riemann solvers perform almost identically for RP1--RP4.
RP5 is more extreme, involving a large density and pressure discontinuity, and it reveals the difference between the Riemann solvers.
The Roe Riemann solver performs the worst out of the three.
The Bassi Riemann solver, whose performance corresponds with the graphs in the original study \cite[Figure 6]{Bassi2018}, is not the best performing either.
This is surprising as the numerical flux should be exact when the contact wave is bounded by the acoustic waves which, indeed, was never violated in this test.
The Osher Riemann solver performs the best.
The results suggest that the Roe Riemann solver should be avoided and, as the validity of the assumptions underpinning the Bassi Riemann solver cannot be guaranteed, the Osher Riemann solver provides a robust alternative.

\subsection{Hydrostatic pressure}
To assess the well-balanced properties of the different pressure gradient calculations, a simple 1D hydrostatic problem was used as a benchmark test, as in Qian et al.\ \cite{Qian2006}.
The test involves a column of air ($\rho = 1$) overlaying a column of water ($\rho = 1000$) under the influence of gravity.
The initial conditions are $\mathbf{u} = 0$ and $p = 0$ everywhere, and the density and velocity values should not change but the pressure should settle to hydrostatic.

As the implementation is 2D, a mesh was created with 20 cells in the $x$-direction and one cell in the $y$-direction, with all four boundaries modelled as walls, and gravity of $\mathbf{g} = (9.81, 0)$.
For pseudo-time, which is explicit, three Runge-Kutta stages and a Courant number of $0.98$ were used with local time stepping.
The simulation was run for one real-time step of size $\Delta t = 0.05$ until the residuals for $\rho u$ were less than $1 \times 10^{-5}$. 
The Osher Riemann solver was used with the minmod limiter for density and velocity in all simulations.
The value of the artificial compressibility coefficient was set to $\beta = 2000$---a smaller value led to instability in the split simulations, a problem acknowledged by Kelecy and Pletcher \cite{Kelecy1997}.
The only difference between the simulations was the pressure gradient calculation.

\begin{figure}
    \centering
    \includegraphics[width = \textwidth]{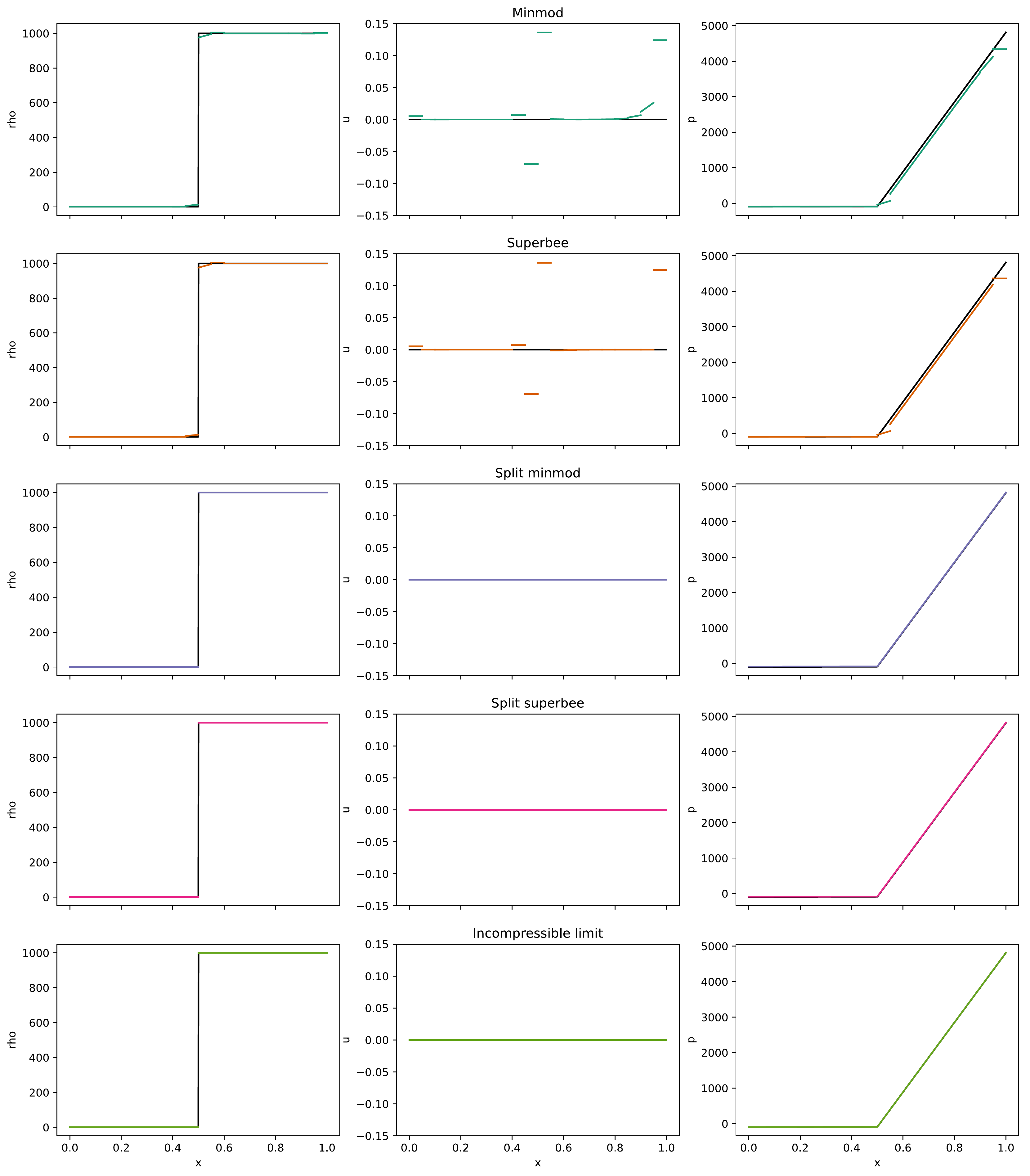}
    \caption{Comparison of pressure gradient calculations in 1D hydrostatic test case.}
    \label{fig:hydrostatic}
\end{figure}

Figure \ref{fig:hydrostatic} shows that limiters with no special treatment do not converge to the correct solution.
The anomalies near the boundaries can be attributed to the reversed gravity in the ghost cells not having any effect for standard limiters, as discussed in Section \ref{sec:boundary}.
However, the spurious velocities near the free surface are precisely due to standard limiters including information from the other side of the free surface, as discussed in Section \ref{sec:gradp}.
In both cases, incorrect pressure gradients lead to incorrect densities or velocities, which is important to remember for the next benchmark test.

In contrast, the two split limiters and the new incompressible limit method converge to the correct solution.
This demonstrates that special treatment is required to ensure a balance with the gravity source term, thereby eradicating spurious densities and velocities.

\subsection{Dam breaks}
In dam breaks, a column of water is initially at rest behind an imaginary, infinitesimally narrow dam.
The dam instantaneously disappears, and the column of water is free to flow.
Dam break simulations are useful for testing the robustness of a free-surface method under extreme conditions, and their widespread use means that experimental data is available for validation.

In this study, we simulated a dam break in a small domain to examine pressure gradient calculations in detail, and then a much larger domain to test the method against the full range of Martin and Moyce's \cite{Martin1952} experimental data.
In both cases, the initial conditions involved a high density ratio, $\rho = 1$ for air and $\rho = 1000$ for water, with $\mathbf{u} = 0$ everywhere and, since the initial condition for pressure is somewhat arbitrary \cite{Kelecy1997}, $p$ is set to zero everywhere initially.
All four boundaries were modelled as walls, but this should not affect comparisons with the experimental data as water is much denser than the air \cite{Bhat2019,Kelecy1997}.
Gravity was set to be $\mathbf{g} = (0,-9.81)$.

\subsubsection{Small domain (non-hydrostatic pressure)}
The previous benchmark showed that split limiters are well-balanced under hydrostatic pressure, unlike standard limiters.
However, recall from Section \ref{sec:gradp} that pressure gradient discontinuities occur even in the absence of gravity.
While split limiters do not address this problem, the new incompressible limit method deals with it automatically.
This was explored with the first real-time step of a dam break.

A mesh was created with 20 cells in both directions.
The Osher Riemann solver was used with the minmod limiter for density and velocity in all simulations.
For pseudo-time, which is explicit, three Runge-Kutta stages and a Courant number of $0.48$ were used with local time stepping.
The simulations were run for one real-time step of size $\Delta t = 0.01$ until the residuals for $\rho u$ and $\rho v$ were smaller than $1 \times 10^{-3}$.
The value of the artificial compressibility coefficient was set to $\beta = 1100$ as smaller values led to instability. 
The only difference between the simulations was the pressure gradient calculation.

\begin{figure}
    \centering
    \includegraphics[width = \textwidth]{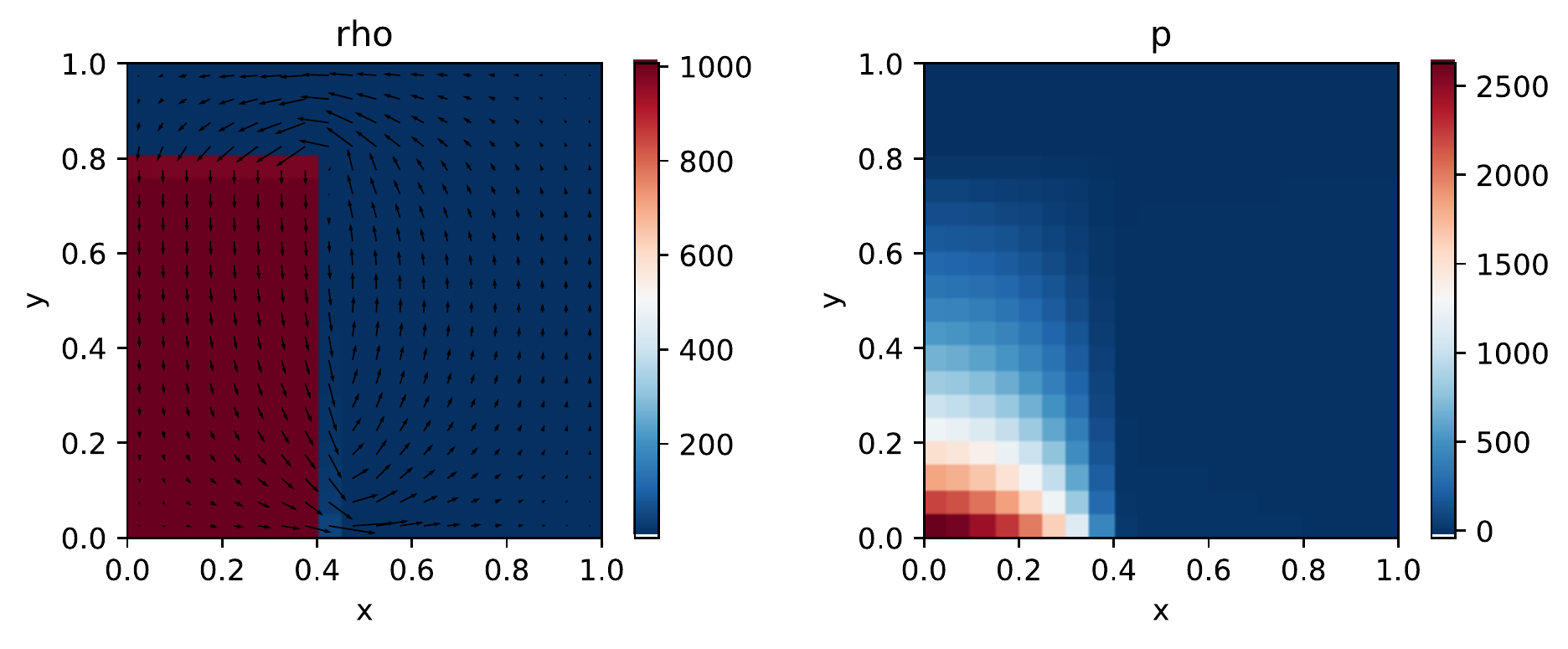}
    \caption{First real-time step of dam break (small domain) with new incompressible limit pressure gradient calculation.}
    \label{fig:dam1_2D}
\end{figure}

\begin{figure}
    \centering
    \includegraphics[width = \textwidth]{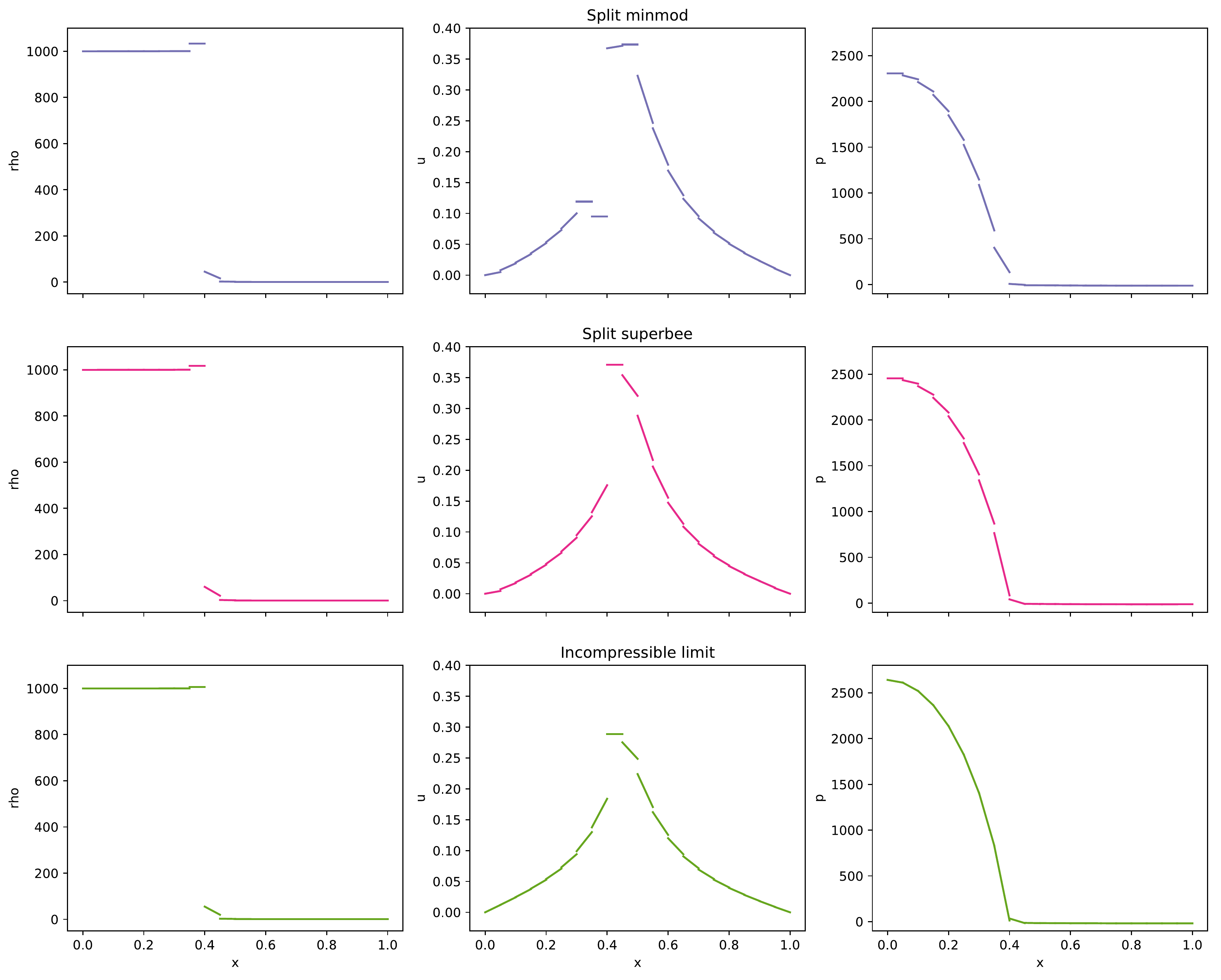}
    \caption{Comparison of pressure gradient calculations in first real-time step of dam break (small domain), along the lowest cross-section parallel to the $x$-axis, which is not the direction of gravity.}
    \label{fig:dam1_1D}
\end{figure}

Figure \ref{fig:dam1_2D} shows the 2D fields in the first real-time step, and Figure \ref{fig:dam1_1D} a cross-section along the lowest row of cells.
As gravity does not act in the $x$-direction, the split and unsplit slope limiter calculations are the same in this direction.
However, the split results are shown here because the unsplit results will be impacted by gravity in the $y$-direction, as illustrated in Figure \ref{fig:hydrostatic} from the previous benchmark.

Consider Figure \ref{fig:dam1_1D} and the pressure gradient just to the left of the density discontinuity at $x = 0.4$.
The split limiters underestimate its magnitude, especially minmod.
This is because information is taken from both sides of the free surface, but there should be a discontinuity defined by the dynamic boundary condition $[\nabla p/\rho] = 0$.
The new incompressible limit method performs better than the split limiters.
In fact, it performs surprisingly well considering that it is not based on the value of pressure, only the other variables, TVD gradients, and source terms.

Now recall from the hydrostatic benchmark test that, the better the pressure gradient calculation at capturing the pressure gradient discontinuity, the smaller the errors in the density and velocity.
Consider Figure \ref{fig:dam1_1D} and the density just to the left of the density discontinuity at $x = 0.4$. All three simulations seem to have converged to a solution with varying degrees of compression.
While this compression resembles Gibbs-type oscillations in the density field, Figure \ref{fig:dam1_1D} clearly shows that the minmod density limiter has fully limited the gradient to zero, and so the density limiter is not causing the problem.
As the only difference between the simulations is the pressure gradient calculation, we must conclude that this is what is causing the problem.
The better the method at capturing the pressure gradient discontinuity, the smaller the error.
The standard limiters, due to their inability to capture the pressure gradient discontinuity, cause the largest errors.
Even though the correct solution for the velocity field is unknown, we might similarly conclude that the pressure gradient is also causing the differences in the velocity values.

\subsubsection{Large domain (experimental data)}
So far, the benchmark tests have not gone further than one real-time step.
In this benchmark, we recreated Martin and Moyce's \cite{Martin1952} dam break experiments for square columns with dimension $a = 2 \frac14$ inches $= 0.05715$ metres and $a = 4 \frac12$ inches $= 0.1143$ metres.
Kelecy and Pletcher \cite{Kelecy1997} also used these data for the same equations, and Bhat and Mandal \cite{Bhat2019} used them for artificial compressibility coupled with VOF.
However, both studies had a domain too small to cover the full temporal range of data, with Kelecy and Pletcher's \cite{Kelecy1997} domain having width $5a$ and height $1.25a$.
Here, we used a domain of width $15a$ and height $1.25a$, divided into 240 cells in the $x$-direction and 20 cells in the $y$-direction.
This allowed the surge to propagate unimpeded until the end of the experiment.

The other parameters included two pseudo-time Runge-Kutta stages, an artificial compressibility coefficient of $\beta = 1100$, and convergence tolerances of 0.01 for $\rho$, $\rho u$, $\rho v$, and $p$.
Both the pseudo-time and real-time Courant numbers were set to 0.45.
The real-time Courant number was set this low not for stability (the real-time derivative is treated point-implicitly) but because it led to better convergence in pseudo-time.
In all cases, the minmod slope limiter was used for density and velocity, but the Riemann solver and pressure gradient calculations were changed.
This was to determine if the variability experienced in the previous benchmarks is important in long simulations.

\begin{figure}
    \centering
    \includegraphics[width = \textwidth]{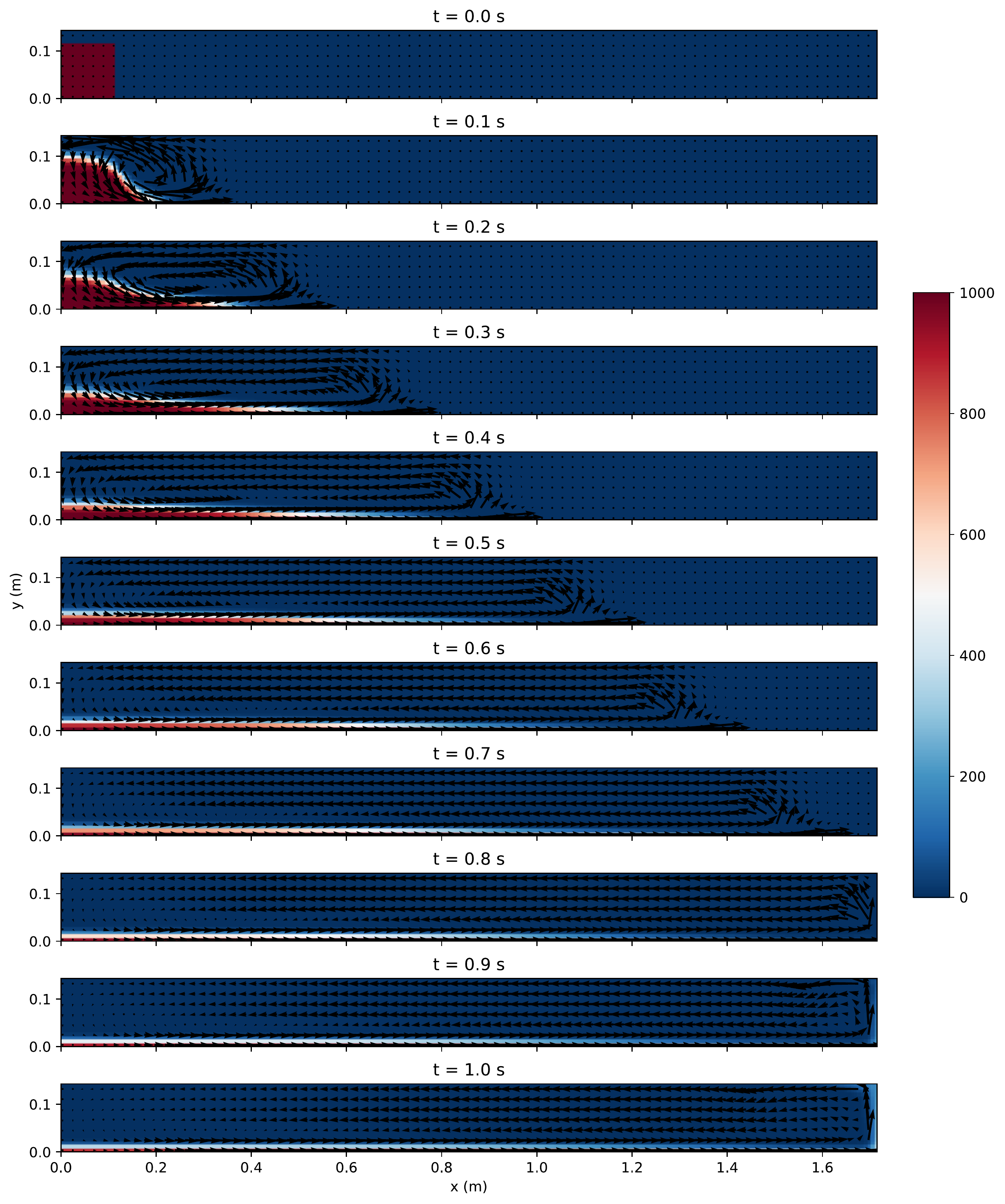}
    \caption{Density in the dam break (large domain) with length scale $a =$ 4 1/2 inches, the Bassi Riemann solver, and the incompressible limit pressure gradient calculation. Velocity shown once every nine cells for clarity.}
    \label{fig:dam2_2D}
\end{figure}

Figure \ref{fig:dam2_2D} shows real-time snapshots of one simulation.
They are similar to the results in \cite{Bhat2019,Kelecy1997}.
In particular, the difference between the density of the water and the density of the air induces a large horizontal pressure gradient, which causes a surge front to propagate out from the column of water, and a vortex to appear around the free surface.
These snapshots explicitly show the density field, rather than a contour to represent the free surface as in \cite{Bhat2019,Kelecy1997}, meaning that the numerical diffusion is not hidden.
Indeed, the transition between water and air straddles several cells, and there is much diffusion towards the end of the simulation when the surge front hits the far wall.
This diffusion could be addressed with the slope modification method \cite{Pan2000} or an interface compression term \cite{Bhat2019a}, which have both been implemented for artificial compressibility coupled with VOF previously.

\begin{figure}
    \centering
    \includegraphics[width = \textwidth]{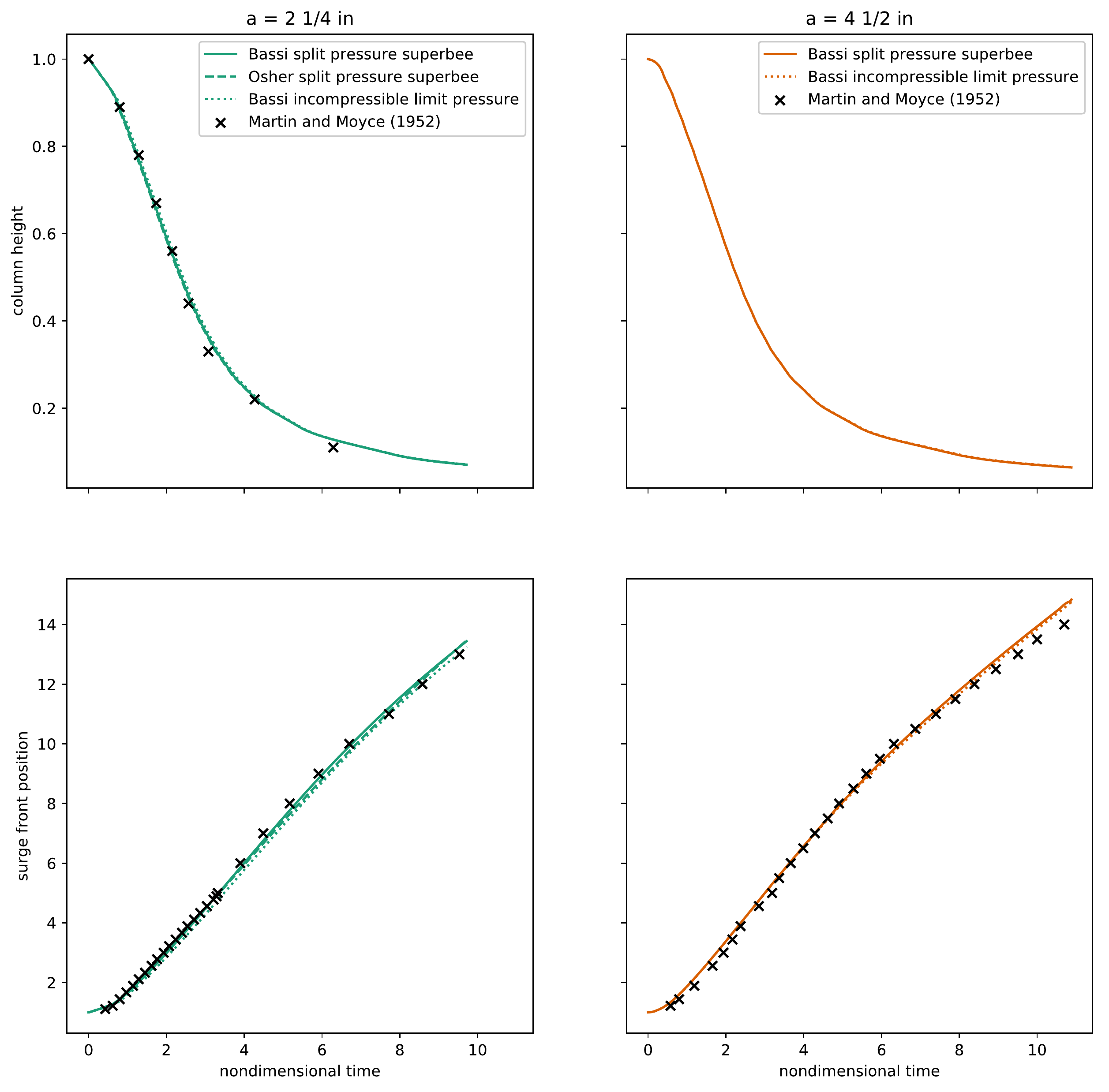}
    \caption{Comparison of dam break (large domain) simulations with experimental data from Martin and Moyce \cite{Martin1952}. Note that column height measurements do not exist for $a =$ 4 1/2 in.}
    \label{fig:dam2_exp}
\end{figure}

Figure \ref{fig:dam2_exp} shows the post-processed results, where the column height and surge front position have been normalised by dividing by $a$, and the time multiplied by $\sqrt{g/a}$.
This allows the two experiments to be compared even though they are at different scales.
The Riemann solver and pressure gradient calculation do not have much of an effect on the results, especially for the $a = 4 \frac12$ inches case, and the simulated values are close to the measured values.
There are some differences between the simulations and observations, particularly in the surge front position towards the end of the $a = 4 \frac12$ inches case, but it is impossible to know if these are due to the model or due to the data, which were collected 70 years ago.
In any case, the results show that the method performs well under this extreme scenario, and the general outputs are not very sensitive to the Riemann solver or pressure gradient calculation.
It is interesting to note that at no point was the wave ordering assumption behind the Bassi Riemann solver violated, which suggests it may also be acceptable for general use alongside the Osher Riemann solver.
However, further investigations are needed to find out when the wave ordering assumption is broken, as acknowledged by Bassi et al.\ \cite[Appendix B4]{Bassi2018}.

\section{Conclusion}
Only a few researchers \cite{Kelecy1997,Qian2006,Bassi2018} have investigated the variable-density artificial-compressibility equations previously, with their developments spaced out over several decades.
This paper presented a Godunov-type scheme, taking advantage of recent advances in the wider field by applying a more robust Riemann solver \cite{Dumbser2011} and a more easily parallelisable time discretisation \cite{Loppi2019,Vermeire2019} to these equations for the first time.
We also developed a new method for calculating the pressure gradient in the MUSCL reconstruction step.
By using the information already available from the TVD gradients and source terms, the new method automatically captures the pressure gradient discontinuity at the free surface.
This pressure gradient method is applicable to any scheme with dual time stepping.

Benchmark tests presented in this paper demonstrated the capabilities of the Godunov-type scheme.
First, a battery of five Riemann problems showed that it can capture discontinuities in the flow, and that the Osher Riemann solver provides a robust alternative to the Bassi Riemann solver.
The Osher Riemann solver may be needed as the Bassi Riemann solver is only exact for certain wave configurations.
Second, the new pressure gradient method has been shown to be well-balanced under hydrostatic cases, like split limiters.
Third, the new pressure gradient method has been shown to capture non-hydrostatic density discontinuities, unlike split limiters.
This suggests that the new pressure gradient method is preferable to existing options.
Finally, Martin and Moyce's \cite{Martin1952} dam break experiments were successfully recreated.
In particular, while Kelecy and Pletcher \cite{Kelecy1997} only used a temporal subset of this experimental data, here an extended domain allowed the full propagation of the surge front to be assessed.
This shows that the method is suitable for practical applications.

Moving forwards, current work is focused on implementing the scheme for unstructured 3D meshes in parallel.
This is largely a software development task as the method outlined in this paper is general apart from the standard and split slope limiters.
The more general implementation will not only harness the ability of the finite volume method to deal with arbitrary cell shapes, but also the ability of the explicit pseudo-time stepping to be easily parallelised.
This will allow the method to be applied in practical free-surface simulations, where there can be millions of finite volume cells in an assortment of shapes and sizes.

\section*{Acknowledgements}
This research was funded by the Engineering and Physical Sciences Research Council grant number EP/R51309X/1.

\bibliography{mybibfile}

\end{document}